\documentclass{article}
\usepackage{cite}
\usepackage{amsmath,amssymb,amsfonts}
\usepackage{graphicx}
\usepackage{textcomp}
\usepackage{xcolor}
\usepackage{booktabs}
\usepackage{multirow}
\usepackage{algorithm}
\usepackage{algorithmic}

\date{}

\begin{document}

\title{Corrigendum and Supplement to \texttt{"}Improve Language Modelling for Code Completion through Learning General Token Repetition of Source Code (with Optimized Memory)\texttt{"}
}
\author{Yixiao Yang \\
School of Software, Tsinghua University \\
yangyixiaofirst@163.com
}

\maketitle

\begin{abstract}
This paper is written because I receive several inquiry emails saying it is hard to achieve good results when applying token repetition learning techniques. If REP \cite{seke19-rep} (proposed by me) or Pointer-Mixture \cite{pointer-mix} (proposed by Jian Li) is directly applied to source code to decide all token repetitions, the model performance will decrease sharply. As we use pre-order traversal to traverse the Abstract Syntax Tree (AST) to generate token sequence, tokens corresponding to AST grammar are ignored when learning token repetition. 
For non-grammar tokens, there are many kinds: strings, chars, numbers and identifiers. 
For each kind of tokens, we try to learn its repetition pattern and find that only identifiers have the property of token repetition. 
For identifiers, there are also many kinds such as variables, package names, method names, simple types, qualified types or qualified names. 
Actually, some kinds of identifiers such as package names, method names, qualified names
or qualified types are unlikely to be repeated. Thus, we ignore these kinds of identifiers that are unlikely to be repeated when learning token repetition. 
This step is crucial and this important implementation trick is not clearly presented in the paper because we think it is trivial and too many details may bother readers. 
We offer the GitHub address of our model in our conference paper and readers can check the description and implementation in that repository. 
Thus, in this paper, we supplement the important implementation optimization details for REP \cite{seke19-rep} and the experiment \cite{ijseke19-rep} compared with Pointer-Mixture. 
\end{abstract}

\section{Corrected REP model Details}
\subsection{The tokens which REP model considers} 
This section will show very detailed implementation optimizations including detailed usage of Eclipse JDT classes and some options in calculating accuracy. This optimization is implemented by the second coauthor of the paper. He thinks the details may confuse readers which is not familiar with Eclipse JDT. But it is our fault that we do not mention the important implementation optimizations in the paper. 
The identifiers in Java AST generated by Eclipse JDT is the leaf node with type $org.eclipse.jdt.core.dom.SimpleName$ (abbreviated as $SimpleName$). 
In paper \cite{ijseke19-rep}, we point out that we split leaf node into two tokens (node type as one token and node content as one token) to check whether the syntax is predicted correctly. We predict the content of leaf node based on the type of leaf node. 
This step has no impact on the traditional language model, but it has a great impact on token repetition learning model such as REP and Point-Mixture. 
Based on this step, we try to use REP model to decide whether the content of a leaf node with type $SimpleName$ should be the previously existed token or not. 
However, considering all leaf nodes with type $SimpleName$ leads to low model performance, we must filter out the kinds of $SimpleName$ nodes which are unlikely to be repeated. 
This filtering step is crucial because if the number of unrepeatable tokens is far more than the number of tokens that will be repeated, the classifier will assume that all tokens will not be repeated. The REP model will degenerate into the traditional language model. Now, we will describe in details what kind of  $SimpleName$ node will be considered by REP model. 

\begin{table}[htbp]
\centering
\caption{Filter Conditions for Node}
\label{filt-cond}
\begin{tabular}{|c|c|p{4cm}|}
\hline
Node Type  & Parent Node Type          & Extra Node Condition \\ \hline
SimpleName & ContinueStatement         & null                 \\ \hline
SimpleName & SimpleType                & null                 \\ \hline
SimpleName & TypeParameter             & null                 \\ \hline
SimpleName & MarkerAnnotation          & null                 \\ \hline
SimpleName & NormalAnnotation          & null                 \\ \hline
SimpleName & MemberValuePair           & null                 \\ \hline
SimpleName & QualifiedType             & null                 \\ \hline
SimpleName & QualifiedName             & null                 \\ \hline
SimpleName & MethodDeclaration         & null                 \\ \hline
SimpleName & LabeledStatement          & null                 \\ \hline
SimpleName & BreakStatement            & null                 \\ \hline
SimpleName & ExpressionMethodReference & null                 \\ \hline
SimpleName & SwitchCase                & null                 \\ \hline
SimpleName & MethodInvocation          & Node is method name  \\ \hline
SimpleName & SuperConstructorInvocation     & Node is super class  \\ \hline
SimpleName & SuperMethodInvocation          & Node is method name or super class \\ \hline
\end{tabular}
\end{table}

We use the rules in Table \ref{filt-cond} to filter out $SimpleName$ nodes which are unlikely to be repeated. 
Note that the $SimpleName$ nodes with parent type $LabeledStatement$, $BreakStatement$, $ContinueStatement$ actually are highly repeated.  However, these nodes correspond to `go-to' semantics in Java language, the proportion of these nodes is too small, and an isolated REP model should be used to learn the token repetition of these `go-to' related nodes. Thus, we filter out these nodes here. 
Even we filter out these nodes, the remaining $SimpleName$ nodes still make up 20\% of the total nodes. 
If the type of a node and the type of its parent node match any row of data in the table, that node will be filtered. For a $SimpleName$ node, if its parent node type is $MethodInvocation$ or $SuperMethodInvocation$, it will be filtered if it meets the $Extra Node Condition$ in the table. 
In our experiment, method name only has a very low probability of being repeated, so it needs to be removed. Thus, if the parent node type of a $SimpleName$ node is $MethodInvocation$ and that $SimpleName$ node represents the name of the invoked method, it will be filtered. Similarly, in some cases, people will call the constructor of the specified parent class, if the parent node type of a $SimpleName$ node is $SuperConstructorInvocation$ and that $SimpleName$ node represents the specified class, that node will be filtered. Note that we have conducted experiments to ensure that the filtered nodes have very low probability of being repeated. We only use simple syntax information to do this filtering. 

\textbf{Cared Node}: after filtering, we give the remaining $SimpleName$ nodes a name: \textbf{Cared Node}. REP model uses syntax to judge whether the node is a \textbf{Cared Node} based on the type of the node. According to the node type on AST, if the node being code completed is a \textbf{Cared Node}, then REP model begins to predict its content. 
Of course, REP model only considers previously existed \textbf{Cared Nodes} in context. 

\begin{figure}[htbp]
\centering
\includegraphics[width=0.98\linewidth]{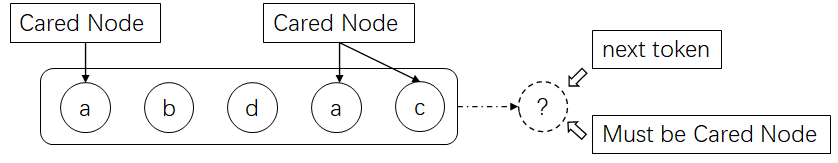}
\caption{Original Token Sequence}
\label{fig:origin-real}
\end{figure}
\begin{figure}[htbp]
\centering
\includegraphics[width=0.74\linewidth]{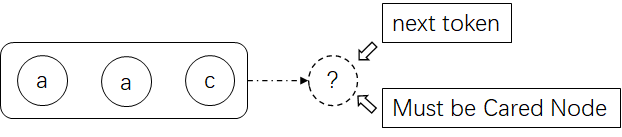}
\caption{Tokens Used by REP}
\label{fig:rep-real}
\end{figure}

For example, Figure \ref{fig:origin-real} shows a token sequence. 
As illustrated in that sequence, token $b$ and token $d$ are not \textbf{Cared Nodes}. 
As REP model only considers \textbf{Cared Nodes}, for REP model, token $b$ and token $d$ should be deleted. 
Figure \ref{fig:rep-real} shows the context which REP model actually uses. 

\subsection{REP model only considers a fixed-length context}
\begin{figure}[htbp]
\centering
\includegraphics[width=0.74\linewidth]{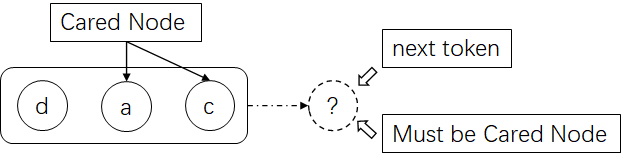}
\caption{Original Context}
\label{fig:origin-ctx}
\end{figure}

For the position to be code-completed, REP model only considers \textbf{Cared Nodes} in the previous $m$ tokens. 
The previous $m$ tokens are taken as context. 
The $m$ is taken as context length. 
For example, if the original token sequence is shown in Figure \ref{fig:origin-real}. If we only consider previous 3 tokens as context. Then the original context is shown in Figure \ref{fig:origin-ctx}. 

\begin{figure}[htbp]
\centering
\includegraphics[width=0.68\linewidth]{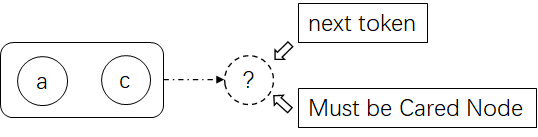}
\caption{Context used by REP}
\label{fig:rep-ctx}
\end{figure}

Here $m$ is 3. As REP only considers \textbf{Cared Nodes}, in Figure \ref{fig:origin-ctx}, token $d$ is not cared, thus, REP model removes token $d$ and only considers token $a$ and token $c$. When $m$ is 3, the context which is considered by REP is shown in Figure \ref{fig:rep-ctx}. The whole idea is very simple. The $m$ is usually set to a small value, for example, 25 or 50 meaning that we only consider 25 previous tokens in learning token repetition. Here, we must correct the setting in paper \cite{ijseke19-rep}: we say we can at most use 600 previous tokens as context. Actually, we use a small number of previous tokens as context. 

\subsection{Corrected REP model algorithm}
The LSTM generates $(cell, h)$ for each token. We use $h_0$, $h_1$..., $h_m$ to denote the $h$ generated by LSTM model for each token in context used by REP. 
\begin{figure}[htbp]
\centering
\includegraphics[width=0.88\linewidth]{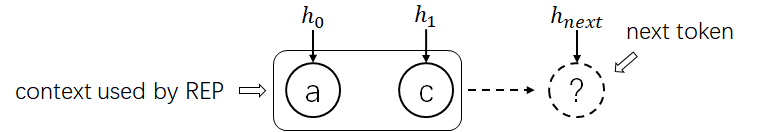}
\caption{Context information used by REP}
\label{fig:rep-ctx-h}
\end{figure}

As shown in Figure \ref{fig:rep-ctx-h}, the $h$ corresponding to tokens in context is denoted as $h_0$, $h_1$, ..., $h_m$. The $h$ corresponding to the token being predicted is denoted as $h_{next}$. 
The probability that $token_{next}$ should be the repetition of $token_k$ in context is computed by:
\begin{equation}
P(k,next) = \frac{e^{h_k^{T} \ W \ h_{next}}}{Z}
\label{eq-edge}
\end{equation}

In Equation \ref{eq-edge}, $Z$ is the normalization factor computed by:
\begin{equation}
Z = \sum_{k=0}^{m} e^{h_k^{T} \ W \ h_{next}}
\label{eq-edge-z}
\end{equation}

In Equation \ref{eq-edge} and \ref{eq-edge-z}, $W$ is the model parameter, $h_k^{T}$ is the transposition of $h_k$. In training phase, if $token_{next}$ is really the repetition of $token_k$ in context, $P(k,next)$ should be maximized. In paper \cite{seke19-rep}, we forget to add the base $e$ in the above equation, this is a mistake and we correct that mistake here. 

To decide $token_{next}$ should be the repetition of some previously existed token or not, we compute $P(token_n\ is\ repeated)$. 
We use symbol $mk$ to denote the $k$th token in context which achieves the maximum probability among P(0,next), P(1,next), ... P(m,next). 
\begin{equation}
mk = \mathop{\arg\max}_{k}\ P(k,next)
\label{eq-max}
\end{equation}

Then $h_{mk}$ is the $h$ for $mk$ th token in context which makes $P(mk,next)$ the highest, $P(token_n\ is\ repeated)$ can be computed as follows:
\begin{equation}
P(token_n\ is\ repeated) =  \frac{e^{h_{mk}^{T}\ V_1\ h_{next}}}{e^{h_{mk}^{T}\ V_1\ h_{next}} + e^{h_{mk}^{T}\ V_2\ h_{next}}}
\label{eq-decision}
\end{equation}

In paper \cite{seke19-rep}, we forget to add the base $e$ in the above equation, this is a mistake again and we correct that mistake here. 
Actually, when training and testing, we use softmax cross entropy to optimize the value. Please see \cite{report-yyx} for implementation details. 

\subsection{Corrected advanced REP model algorithm}
Actually, we can use different isolated REP models to learn token repetition for different kinds of tokens. 
For example, for types related to Java templates (generic class), we can use one REP model to specifically learn the token repetition, for identifiers related to `go-to' syntax in Java, we can use another different REP model to learn token repetition. 
This optimization is implemented by the second author of the paper. Because the second author was in the process of graduation, there was a little mistake in our communication. I didn't mention this important optimization in my paper, so I hereby add these implementation details here so as not to confuse readers. 

\subsection{More advanced REP model algorithm} 
Actually, we can only take Java variables in context into consideration to further improve the accuracy. We can remove those \textbf{Cared Nodes} which are not variables. This step is easily achieved as we can use Eclipse JDT to know which identifier is Java variable or not. 
Here, we provide the details about how to identify all variables in source code using eclipse JDT. If there are other better ways to identify all variables in source code, please ignore the following content. 
When pre-processing, we use eclipse JDT to identify every variable in a function. In details, the eclipse JDT provides a technique named as \emph{ResolveBinding}. For every ASTNode which type is $SimpleName$, we invoke \emph{resolveBinding} method provided by eclipse JDT, if the binding is successfully resolved and the binding type is Variable, we think this ASTNode is a variable. This step is introduced in my Phd thesis, not clearly in paper \cite{ijseke19-rep}. 

\section{Corrected Experiments}
\subsection{Corrected accuracy computation method}
For each function in test set, we start to predict token from start to end. As code-structure-tokens are not predicted by REP model. Taking them into consideration may confuse readers. However, we still take most of them into consideration in our paper. Here, we also give results which do not take irrelevant tokens into consideration to show readers about the very strong ability of token repetition learning. 
Note that, when computing accuracy, some works do not count UNK tokens or some meaningless grammar tokens. In our work, although we split a leaf node into two tokens (node type and node content), we think predicting node content correctly is most important. Thus, when predicting leaf node, we compute the accuracy of predicting node content not node type. When predicting node content, we assume that the node type is already predicted but not compute that prediction accuracy. 
We consider top-k accuracy as the evaluation metrics. The entropy and the mrr are no longer taken into consideration. In this corrected version, we still use training set, validation set and test set. The proportion is 60\%, 20\%, 20\% (slightly different from the paper). 

\subsection{Corrected experimental setting}
The whole training procedure will stop if the top-1 accuracy on validation set does not exceed the maximum for \textbf{10} epochs. The traditional language model and REP model are trained separately. The REP model directly uses the results of the fully-trained traditional language model. This step needs some engineering works please check GitHub address \cite{report-yyx} for details. All initial values for all token embedding parameters are randomly selected between -1.0 and 1.0 (we use uniform\_random initializer in Tensorflow). 
All other parameters such as parameters in LSTM or token repetition are set to 0. This may have some exceptions, please check our implementation \cite{report-yyx}. This setting can maximize the prediction effect for both LSTM and REP. 
The gradient is clipped between $-10^6$ and $10^6$ (we try not to clip gradients). 
We mark 1000 least frequently appeared tokens in training set are marked as UNK. 

\subsection{Corrected experimental results}
For project Log4J, the context length is set to 25 which means REP and Atten-Ptr (token repetition learning in Pointer-Mixture) model only consider cared nodes in previous 25 tokens. 
Table \ref{table:accuracy-small} shows the accuracy. 
As can be seen, learning token repetition can greatly improve the prediction accuracy of tokens especially for unseen tokens. 
\begin{table}[htbp]
\centering
\caption{Accuracy on Log4J}
\label{table:accuracy-small}
\begin{tabular}{|c|c|c|c|c|c|}
\hline
\multicolumn{6}{|c|}{test set}                                                                    \\ \hline
all nodes          & top1          & top3          & top6          & top10         & total number \\ \hline
LSTM               & 46.7          & 58.7          & 65.0          & 68.4          & 10543        \\ \hline
Atten-Ptr          & 51.8          & 64.8          & 71.1          & 74.1          & 10543        \\ \hline
REP                & \textbf{52.0} & \textbf{65.0} & \textbf{71.3} & \textbf{74.3} & 10543        \\ \hline
cared nodes        & top1          & top3          & top6          & top10         & total number \\ \hline
LSTM               & 20.8          & 29.2          & 33.4          & 36.4          & 2227         \\ \hline
Atten-Ptr          & 44.8          & 58.0          & 62.4          & 63.3          & 2227         \\ \hline
REP                & \textbf{45.6} & \textbf{58.9} & \textbf{63.0} & \textbf{64.2} & 2227         \\ \hline
unseen cared nodes & top1          & top3          & top6          & top10         & total number \\ \hline
LSTM               & 0.0           & 0.0           & 0.0           & 0.0           & 681          \\ \hline
Atten-Ptr          & \textbf{27.7} & 37.3          & 39.9          & 40.2          & 681          \\ \hline
REP                & 27.0          & \textbf{38.3} & \textbf{40.8} & \textbf{41.0} & 681          \\ \hline
\multicolumn{6}{|c|}{validation set}                                                              \\ \hline
cared nodes        & top1          & top3          & top6          & top10         & total number \\ \hline
LSTM               & 13.9          & 19.3          & 22.2          & 24.2          & 2646         \\ \hline
Atten-Ptr          & 34.6          & 51.7          & 56.4          & 57.4          & 2646         \\ \hline
REP                & \textbf{37.5} & \textbf{54.4} & \textbf{59.7} & \textbf{60.8} & 2646         \\ \hline
\end{tabular}
\end{table}

\section{Related Work}
The statistical language models have been widely used in capturing patterns of source code to solve the problem of code completion. 
In \cite{DBLP:conf/icse/HindleBSGD12}, source code was parsed into lexical tokens and the n-gram model was applied directly to suggest the next lexical token. In \cite{DBLP:conf/msr/AllamanisS13a}, a large scale experiments was conducted by using n-gram model and a visualization tool was provided to inspect the performance of the language model for the task of code completion. 
In SLAMC \cite{DBLP:conf/sigsoft/NguyenNNN13}, based on basic n-gram model, associating code lexical tokens with roles, data types and topics was one way to improve the prediction accuracy. 
Cacheca \cite{Tu2014On} improved n-gram model by caching the recently encountered tokens in local files to improve the performance of basic n-gram model. 
Decision tree learning was applied to code suggestion, based on this, a decision tree model which integrates the basic n-gram \cite{DBLP:conf/oopsla/RaychevBV16} was proposed for source code. 
The work \cite{poplRaychevBVK16} abstracted source code into DSL
and kept sampling and validating on that specially designed DSL until the good code suggestion was obtained. 
Deep learning techniques such as RNN, LSTM were applied to code generation model \cite{White2015Toward} \cite{dam2016deep} \cite{FSE17} to achieve a higher prediction accuracy. 
The work in \cite{FSE17} confirmed that LSTM significantly
outperforms other models for doing token-level code suggestion.
Given large amount of unstructured code, deep language models such as LSTM or its variants are the state-of-art solutions to the problem of code completion. 
All works described above are trying to solve the general code completion problem in which every token of code should be predicted and completed based on the context in a fixed or changeable length. 
There are also a lot of works paying attention to the API completion problem. 
Common sequences of API calls were captured with per-object n-grams in \cite{DBLP:conf/pldi/RaychevVY14}. 
In \cite{DBLP:conf/icse/NguyenN15}, API usages was trained on graphs. Naive-Bayes was integrated into n-gram model to suggest API patterns. The migrations of API are studied in \cite{Nguyen2017Exploring}. The completion of API full qualified name is studied in \cite{phan2018statistical}. 
On top of general code synthesis problems, API synthesis is also studied in \cite{Nguyen2016T2API,APILearn}. 

\bibliographystyle{IEEEtran}
\bibliography{IEEEabrv,reference.bib}


\end{document}